# Spatially Resolved Conductivity of Rectangular Interconnects considering Surface Scattering- Part I: Physical Modeling

Xinkang Chen, and Sumeet Kumar Gupta, Senior *Member,* IEEE

***Abstract*—Surface scattering in interconnects is usually treated by using Fuchs-Sondheimer (FS) approach. While the FS model offer explicit inclusion of the physical parameters, it lacks spatial dependence of conductivity across the interconnect cross-section. To capture the space-dependency of conductivity, an empirical modeling approach based on "cosh" function has been proposed, but it lacks physical insights. In this work, we present a 2D spatially resolved FS (SRFS) model for rectangular interconnects derived from the Boltzmann transport equations. The proposed SRFS model for surface scattering offers both spatial dependence and explicit relation of conductivity to physical parameters such as mean free path and specularity of electrons (*p*) as well as interconnect geometry. The solution obtained from our SRFS model is exact for diffusive scattering. For specular scattering, we approximate the solution and show that the average conductivity obtained from SRFS shows a good match with the previous models for a general value of *p*. We highlight the importance of physics-based spatially resolved conductivity model by showing the differences in the spatial profiles between the proposed physical approach and the previous empirical approach. In Part II of this work, we build upon the SRFS approach to propose a compact model for spatially-resolved conductivity accounting for surface scattering in rectangular interconnects.**

## I. INTRODUCTION

Technology scaling has been amongst the most important drivers for the advancement of electronic devices. However, several issues are coming to the fore that impair the effectiveness of scaling, amongst which, interconnects scaling is becoming a major bottleneck [1]. The challenges in the interconnect scaling are manifold. Reducing the interconnect width increases their resistance due to lower cross-sectional area. But, more importantly, the resistivity ($\rho_W$) also increases with width scaling due to increase in the sidewall scattering [2]. This issue is further aggravated in standard interconnects that utilize thin barrier/liner layers to mitigate the electromigration in copper (Cu) [3], [4]. The barrier/liner layers do not scale proportionally with technology scaling, leading to the active conduction area that is lower than the interconnect footprint. This further increases the sidewall scattering. To counter this, alternate designs including novel interconnects materials [5], [6], [7] and structures [8] are being explored.

This work was supported , in part, by the NEW materials for LogIc, Memory and InTerconnectS (NEWLIMITS) Center funded by the Semiconductor Research Corporation (SRC)/National Institute of Standards and Technology (NIST) under Award 70NANB17H041. (Corresponding author: Xinkang Chen.)

Xinkang Chen and Sumeet Kumar Gupta are with the School of Electrical and Computer Engineering, Purdue University, West Lafayette, IN 47907 USA (e-mail: chen3030@purdue.edu; guptask@purdue.edu).

To understand the pros and cons of interconnect materials and topologies, modeling their conductivity is highly important. To that end, several models have been proposed. The sidewall scattering is usually treated using the Fuchs-Sondheimer (FS) theory [9]. This approach has been used to model the resistivity of thin film (Fig. 1a) [9], square wires with diffusive scattering (Fig. 1b) [10] and circular wires [11]. Another important scattering mechanism due to grain boundaries is captured using the Mayadas-Shatzkes (MS) theory [12]. With these theories forming the bedrock of conductivity modeling, several works have utilized variants of FS and MS models to estimate the conductivity in scaled technologies [13], [14], [15].

Another important approach for sidewall scattering was proposed in [16], in which the wire resistivity is obtained based on kinetic theory. For wires with square/rectangular cross-section, this approach yields an exact result for perfectly diffusive surface scattering (specularity $p$=0). However, for general $p$ ($p \neq 0$), the exact solution is challenging to obtain for rectangular wires [17]. Thus, for general $p$, resistivity model based on [16] an infinite series expansion was proposed [16].

However, one limitation of most of the existing sidewall scattering models is that the conductivity is not spatially resolved but is an average value across the whole cross section [18]. Modern tapered interconnect structures [19], particularly vias, exhibit non-trivial vertical current paths (Fig. 2a), requiring models that can predict the space-dependent conductivity. For example, the current spreading at Cu-liner or liner/barrier interfaces [15] necessitates the modeling of conductivity variation across the cross-section in these layers.

A notable exception amongst various works on interconnect modeling is [20] that does consider the spatial dependence of conductivity due to sidewall scattering. In this work, the authors

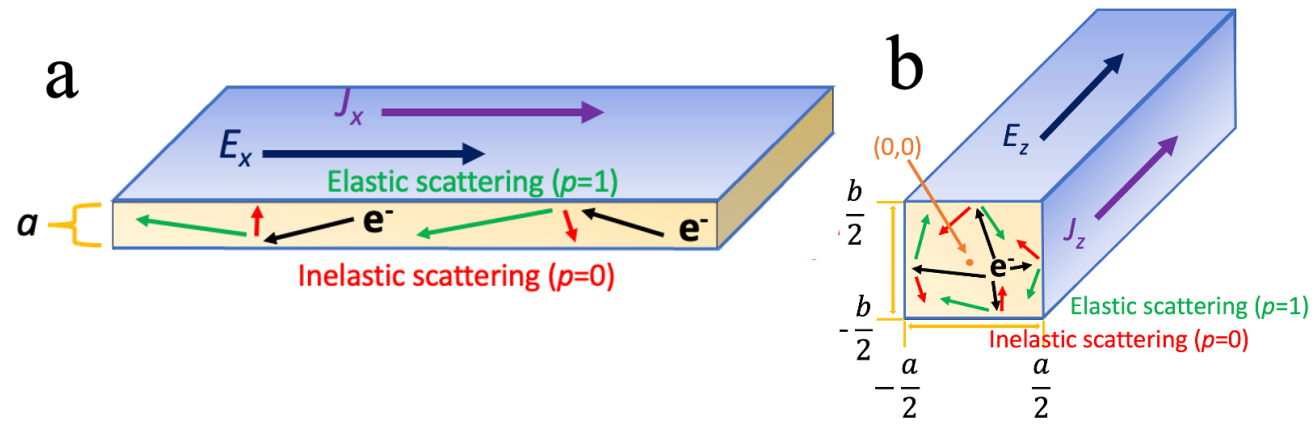

Fig. 1. a) Thin film structure b) rectangular wire structure used for modeling showing fully elastic (specularity $p$=1) or inelastic scattering ($p$=0) when electron hits surfaces

use an empirical approach using a "cosh" function to model the conductivity as a function of location in the interconnect cross-section. However, being an empirical approach, this technique lacks the physical insights that the FS theory provides.

In this two-part paper, we bridge the gap between the FS approach and the "cosh"-based model [20] by obtaining a spatially resolved model for interconnect conductivity while retaining the physical insights. In Part I, we derive the spatial dependence of the conductivity of rectangular/square wires (Fig. 2a) based on the FS theory. For $p$=0, our conductivity model is exact and consistent with the FS theory. For $p\neq 0$, we propose some approximations and present a spatially resolved conductivity model for a general $p$. Based on this, in Part II, we present a circuit-compatible model for interconnect conductivity considering sidewall scattering and accounting for spatial dependence as well as the dependence on physical parameters. The proposed spatially resolved conductivity models (especially the circuit-compatible version) can be integrated in a finite element simulation [15] to obtain the overall interconnect resistance, accounting for non-trivial current transport. Note, that in this process, one can also account for surface roughness along the current direction by cross-sections slices of varying dimensions, as shown in Fig. 2b. However, the proposed model does not explicitly consider surface roughness in the cross-sectional slice (Fig. 2c). It is worth clarifying that the objective of this work is to only propose spatially resolved conductivity model; its deployment in a finite element simulation to predict the overall resistance and the effect of spatially resolved conductivity on interconnect delay is a subject of future research. The main contributions of Part I of this work are as follows:

- We derive an expression for spatially resolved FS (SRFS) conductivity for rectangular interconnects based on Boltzmann transport equation (BTE) for diffusive surface scattering ($p$=0). The model comprises of physical parameters such as electron mean free path ($\lambda_0$) and interconnect width/height.
- Based on this, we propose an approximate solution for spatially resolved conductivity for specular scattering (generic $p$) and validate our approach by comparing the average conductivity with previous works [16], [21].
- We compare the proposed model with the previous "cosh"-based modeling approach [20], highlighting the importance of the physics-based spatial resolution of conductivity and laying the groundwork for Part II of this paper.

## II. Spatially Resolved FS (SRFS) Model for Rectangular Wires

We begin by presenting the derivation of the space-dependence of the surface scattering model for rectangular wires following the FS approach, which we refer to as the spatially resolved FS or SRFS model. We start with purely diffusive scattering i.e. specularity ($p$)=0 (exact solution). We then generalize the discussion for an arbitrary $p$ (between 0 and 1). We assume that the electric field ($E_Z$) and the current flow is along the $z$-direction, and the wire has width and height equal to $a$ and $b$ (along the $x$- and $y$- directions), respectively – Fig. 1b. We consider the origin of the coordinate system to be in the center of the wire cross-section. Note, we neglect explicit surface roughness in the cross-section (similar to FS model); however, one may abstract the effect of surface roughness in specularity, when calibrating the model with experiments [22].

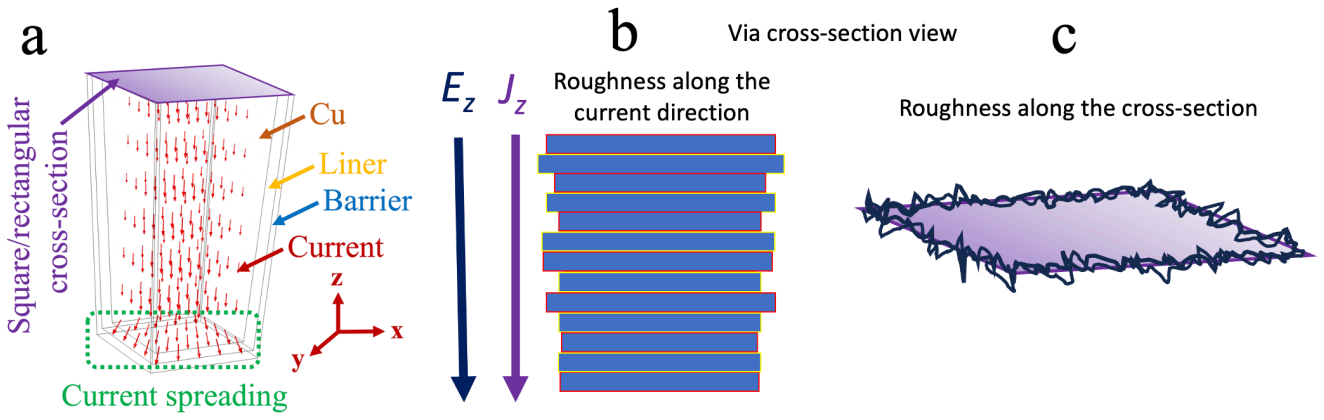

Fig. 2. a) Tapered via structure with copper conductor and barrier/liner, showing non-trivial current transport (such as current spreading effect at the bottom of the via) b) line-edge roughness along the current direction c) line-edge roughness along the cross-section.

### A. Purely Diffusive Surface Scattering (p=0)

The FS approach uses the Boltzmann Transport Equations (BTE) to obtain the conductivity, which involves obtaining the deviation of the electron distribution function from its equilibrium value ($\Delta f$). We follow the approach in [18] accounting for scattering at the four surfaces of the wire to obtain $\Delta f$ for a rectangular wire as

$$\Delta f_{s1,s2} = \frac{q\tau E_z}{m_{eff}}\frac{\partial f_0}{\partial v_z}\left(1 - e^{-\frac{\min\left(\frac{x+(s1)\frac{a}{2}}{v_x},\ \frac{y+(s2)\frac{b}{2}}{v_y}\right)}{\tau}}\right) \quad (1)$$

$s1=sign(v_x)$, $s2=sign(v_y)$, i.e. $s1, s2 \in \{-1, +1\}$.

Here, $s1$ = sign($v_x$), $s2$ = sign($v_y$), where $v_x$, $v_y$ and $v_z$ are the electron velocities in the $x$-, $y$- and $z$-directions. Hence, $s1, s2 \in \{-1, +1\}$. Thus, there are four possible expressions for $\Delta f$ viz. $\Delta f_{+1,+1}$, $\Delta f_{-1,+1}$, $\Delta f_{+1,-1}$, and $\Delta f_{-1,-1}$ that cover the $v_x$-$v_y$ plane. Other terms in (1) are defined as follows: $q$ is the electronic charge, $m_{eff}$ is the effective mass of the electrons, $\tau$ is the relaxation time, and $f_0$ is the equilibrium distribution functions of the electrons. The $\Delta f$ functions are used in the BTE for quasi-free electrons to obtain the current density $J_z$ as

$$J_z(x,y) = -2q\left(\frac{m_{eff}}{h}\right)^3 \sum_{s1,s2} \iiint v_z \Delta f_{s1,s2}\, dv_x dv_y dv_z \quad (2)$$

where $h$ is the Planck's constant.

In the existing FS models, the average conductivity is obtained by dividing (2) by $E_Z$ and averaging along the $x$ and $y$ directions. In order to retain the spatial dependency of the conductivity, the current density ($J_z(x,y)$) is divided by $E_z$ without averaging. Thus, spatially-resolved conductivity ($\sigma(x,y)$) accounting for surface scattering can be obtained as

$$\sigma(x,y) = -\frac{2q^2 {m_{eff}}^2}{h^3} \times \sum_{s1,s2} \iiint \frac{\partial f_0}{\partial v}\frac{{v_z}^2\tau}{v}\zeta\left(\frac{x+(s1)\frac{a}{2}}{\tau v_x}, \frac{y+(s2)\frac{b}{2}}{\tau v_y}\right) dv_x dv_y dv_z \quad (3)$$

Here $\zeta$ is a function that depends on the spatial location ($x$ and $y$) and the electron velocities $v_x$ and $v_y$ and is derived by substituting the $\Delta f_{s1,s2}$ functions of (1) in the expression for $J_z$ in (2). We will derive the explicit form of this function subsequently.

Now, following the FS approach, a degenerate electron gas is assumed, which implies that for any function $g(v)$, the following holds.

$$\int_0^{\infty} g(v)\frac{\partial f_0}{\partial v}dv = -g(\tilde{v}) \tag{4}$$

Here, $\tilde{v}$ is the magnitude of the electron velocity at the Fermi surface. Using this condition in (3), transforming the electron velocities to the spherical coordinate system, and substituting the electron mean free path $(\lambda_0) = \tau\tilde{v}$, we obtain $\sigma(x,y)$ as

$$\begin{aligned}\sigma(x,y) &= \frac{2q^2{m_{eff}}^2}{h^3}\sum_{s1,s2}\int_{\theta=0}^{\pi}\int_{\Phi=\Phi_l}^{\Phi_h}\tau\tilde{v}^3 cos^2\theta sin\theta \\ &\times\zeta\left(\frac{x+(s1)\frac{a}{2}}{\lambda_0 sin\theta cos\Phi},\frac{y+(s2)\frac{b}{2}}{\lambda_0 sin\theta sin\Phi}\right)d\Phi d\theta\end{aligned} \tag{5}$$

Here, $\Phi$ and $\theta$ are the azimuthal and the polar angles, respectively. $\{\Phi_l,\Phi_h\} = \left(0,\frac{\pi}{2}\right),\left(\frac{\pi}{2},\pi\right),\left(\pi,\frac{3\pi}{2}\right),\left(\frac{3\pi}{2},2\pi\right)$ for $\{s1,s2\} = (+1,+1),(-1,+1),(-1,-1),(+1,-1)$ respectively. Now, we substitute the bulk conductivity $(\sigma_0) = \frac{8\pi}{3}\frac{q^2 m_{eff}^2\tilde{v}^2}{h^3}\lambda_0$ in (5). Further, we follow a common practice used to express the FS model in terms of the ratio of the wire width/height and the electron mean free path [18]. For that, we define $\kappa_a = \frac{a}{\lambda_0}$ and $\kappa_b = \frac{b}{\lambda_0}$. We also normalize $x$ and $y$ as $x_n = \frac{x}{a/2}$ and $y_n = \frac{y}{b/2}$. Hence, $x_n$ and $y_n$ range from -1 to 1. Finally, we obtain

$$\begin{aligned}&\sigma(x_n,y_n) = (3/4\pi)\times\sigma_0 \\ &\sum_{s1,s2}\int_{\theta=0}^{\pi}\int_{\Phi=\Phi_l}^{\Phi_h}\zeta\left(\frac{\kappa_a(x_n+s1)}{2sin\theta cos\Phi},\frac{\kappa_b(y_n+s2)}{2sin\theta sin\Phi}\right)cos^2\theta sin\theta d\Phi d\theta\end{aligned} \tag{6}$$

Let us first consider the integral of $\zeta$ with respect to $\Phi$, which we define as $\eta(x_n,y_n,\theta)$. Thus,

$$\sigma(x_n,y_n) = \frac{3}{4\pi}\sigma_0\int_{\theta=0}^{\pi}\eta(x_n,y_n,\theta)cos^2\theta sin\theta d\theta \tag{7}$$

Now to evaluate $\eta$, we consider four cases based on the signs of $v_x$ and $v_y$ (i.e. the values of s1 and s2 - see (1)): (i) $v_x > 0$, $v_y > 0$, for which $\Phi = 0 \rightarrow \frac{\pi}{2}$, and $\Delta f = \Delta f_{+1,+1}$, (ii) $v_x < 0$, $v_y > 0$, ($\Phi = \frac{\pi}{2} \rightarrow \pi$, and $\Delta f = \Delta f_{-1,+1}$), (iii) $v_x > 0$, $v_y < 0$, ($\Phi = \pi \rightarrow \frac{3\pi}{2}$, and $\Delta f = \Delta f_{+1,-1}$), and (iv) $v_x < 0, v_y < 0$, ($\Phi = \frac{3\pi}{2} \rightarrow 2\pi$, and $\Delta f = \Delta f_{-1,-1}$). Thus, $\eta$ can be written as

$$\begin{aligned}&\eta(x_n,y_n,\theta) \\ &= \sum_{s1,s2}\int_{\Phi=\Phi_l}^{\Phi_h}\zeta\left(\frac{\kappa_a(x_n+s1)}{2sin\theta cos\Phi},\frac{\kappa_b(y_n+s2)}{2sin\theta sin\Phi}\right)d\Phi \\ &= \int_{\Phi=0}^{\frac{\pi}{2}}\zeta(\Phi)|_{\Delta f=\Delta f_{+1,+1}}d\Phi + \int_{\Phi=\frac{\pi}{2}}^{\pi}\zeta(\Phi)|_{\Delta f=\Delta f_{-1,+1}}d\Phi \\ &+ \int_{\Phi=\pi}^{\frac{3\pi}{2}}\zeta(\Phi)|_{\Delta f=\Delta f_{-1,-1}}d\Phi + \int_{\Phi=\frac{3\pi}{2}}^{2\pi}\zeta(\Phi)|_{\Delta f=\Delta f_{+1,-1}}d\Phi\end{aligned} \tag{8}$$

We define the four integrals of the right hand side of (8) as $\eta_{+1,+1},\eta_{-1,+1},\eta_{+1,-1},\eta_{-1,-1}$, respectively i.e.

$$\eta(x_n,y_n,\theta) = \eta_{+1,+1} + \eta_{-1,+1} + \eta_{+1,-1} + \eta_{-1,-1} \tag{9}$$

Let us first look at $\eta_{+1,+1}$. It can be observed that the *min* function in (1) dictates that if $\frac{v_y}{v_x} = \frac{vsin\theta sin\Phi}{vsin\theta cos\Phi} = tan\Phi < \frac{y+b/2}{x+a/2} = \frac{b}{a}\left(\frac{1+y_n}{1+x_n}\right)$, $\Delta f_{+1,+1}$ in (1) uses $\left(1-e^{\frac{-\kappa_a(1+x_n)}{2sin\theta cos\Phi}}\right)$, else it uses $\left(1-e^{\frac{-\kappa_b(1+y_n)}{2sin\theta sin\Phi}}\right)$. Thus,

$$\begin{aligned}\eta_{+1,+1} &= \int_0^{tan^{-1}\left(\frac{b}{a}\left(\frac{1+y_n}{1+x_n}\right)\right)}\left(1-e^{\frac{-\kappa_a(1+x_n)}{2sin\theta cos\Phi}}\right)d\Phi \\ &+ \int_{tan^{-1}\left(\frac{b}{a}\left(\frac{1+y_n}{1+x_n}\right)\right)}^{\frac{\pi}{2}}\left(1-e^{\frac{-\kappa_b(1+y_n)}{2sin\theta sin\Phi}}\right)d\Phi\end{aligned} \tag{10}$$

In a similar manner, when we repeat this for other $\eta$ functions and change the limits of integration to $0 \rightarrow \frac{\pi}{2}$, we get the following expressions.

$$\begin{aligned}\eta_{s1,s2} &= \int_0^{tan^{-1}\left(\frac{b}{a}\left(\frac{1+(s2)y_n}{1+(s1)x_n}\right)\right)}\left(1-e^{\frac{-\kappa_a(1+(s1)x_n)}{2sin\theta cos\Phi}}\right)d\Phi \\ &+ \int_{tan^{-1}\left(\frac{b}{a}\left(\frac{1+(s2)y_n}{1+(s1)x_n}\right)\right)}^{\frac{\pi}{2}}\left(1-e^{\frac{-\kappa_b(1+(s2)y_n)}{2sin\theta sin\Phi}}\right)d\Phi\end{aligned} \tag{11}$$

Substituting (11) in (9), we can rewrite the expression for $\eta$ as

$$\begin{aligned}\eta(x_n,y_n,\theta) = \sum_{n,d}\Bigg[&\int_0^{tan^{-1}\left(\frac{bn}{ad}\right)}\left(1-e^{\frac{-\kappa_a\times d}{2sin\theta cos\Phi}}\right)d\Phi \\ &+ \int_{tan^{-1}\left(\frac{bn}{ad}\right)}^{\frac{\pi}{2}}\left(1-e^{\frac{-\kappa_b\times n}{2sin\theta sin\Phi}}\right)d\Phi\Bigg]\end{aligned} \tag{12}$$

where $(n,d) \rightarrow \{(1+y_n,1+x_n),(1-y_n,1+x_n),(1+y_n,1-x_n),(1-y_n,1-x_n)\}$.

Let us call these four points defined by $(n,d)$ as the boundary points. Depending on the spatial location, the ascending order of the four boundary points from 0 to $\frac{\pi}{2}$ can have eight different possibilities. In general, the order of the boundary points may be needed to simplify (12), which, being space-dependent, may make the expressions complicated. However, we solve (12) following an approach which makes the order of the boundary points irrelevant and thus considerably simplifies the final SRFS conductivity model. The details of our approach are provided in Appendix A1.

Substituting the simplified expression of (12) (see A1) in (8), we obtained the final space-dependent conductivity as

$$\sigma(x_n,y_n) = \frac{3}{4\pi}\sigma_0\int_{\theta=0}^{\pi}\eta(x_n,y_n,\theta)cos^2\theta sin\theta d\theta \tag{13}$$

$$\begin{aligned}\eta(x_n,y_n,\theta) &= 2\pi \\ &-2\int_{\Phi=0}^{\frac{\pi}{2}}\left\{\begin{aligned}&e^{\frac{-\kappa_b}{2sin\theta sin\Phi}}\times\cosh\left(\frac{\kappa_b y_n}{2sin\theta sin\Phi}\right) \\ &+e^{\frac{-\kappa_a}{2sin\theta cos\Phi}}\times\cosh\left(\frac{\kappa_a x_n}{2sin\theta cos\Phi}\right)\end{aligned}\right\}d\Phi \\ &-\frac{1}{2}\sum_{n,d}\int_{\Phi=0}^{\frac{\pi}{2}}\left|e^{\frac{-\kappa_a\times d}{2*sin\theta cos\Phi}}-e^{\frac{-\kappa_b\times n}{2*sin\theta sin\Phi}}\right|d\Phi\end{aligned} \tag{14}$$

where $(n,d) \rightarrow \{(1+y_n, 1+x_n), (1-y_n, 1+x_n), (1+y_n, 1-x_n), (1-y_n, 1-x_n)\}$ and $x_n \in [-1,1],\ y_n \in [-1,1]$

*B. Specular Surface Scattering (General p)*

Let us now generalize the SRFS model accounting for specularity ($p$) at the surfaces (which models both diffusive (inelastic) and elastic scattering at the surface). Here, we consider that all the four surfaces have the same value of $p$.

As noted earlier, the exact solution for general $p$ is challenging to obtain for square/rectangular wires [17] (although for thin films and circular cross-sections, the exact solutions have been obtained [10], [11]). Therefore, here, we propose approximate expressions for $\Delta f$ and follow the approach we discussed in the previous sub-section to obtain the spatially resolved conductivity. We adopt the methodology proposed for thin film as described in [18]. In this approach, for a general $p$, it is assumed that $p$ fraction of electrons scatter elastically, resulting in no change in energy, while 1-$p$ fraction of electrons lose their energy completely. The boundary conditions are modified accordingly. Note, for $p$=0, the boundary condition at the surfaces is that $\Delta f$=0 (see (1)). For a general $p$, the boundary conditions are modified as per the process in [18] i.e. at any surface, $\Delta f$ for electrons travelling away from it = $p$ times $\Delta f$ corresponding to electrons travelling towards it (more details in [18]). Utilizing this methodology and assumptions, the approximate expression for $\Delta f$ are given below.

$$\Delta f_{s1,s2} = \frac{q\tau E_z}{m_{eff}} \frac{\partial f_0}{\partial v_z} \times \left( 1-(1-p) \times \max \left\{ \frac{e^{-\frac{x+(s1)\frac{a}{2}}{\tau v_x}}}{1-p\, e^{-a/\tau v_x}}, \frac{e^{-\frac{y+(s2)\frac{b}{2}}{\tau v_y}}}{1-p\, e^{-a/\tau v_y}} \right\} \right) \quad (15)$$

The choice of the approximations are based on the following observations and conditions. First, the solution obtained from the SRFS model is exact for $p$=0, while for $p$=1 the solution is trivial (i.e. spatially uniform conductivity = bulk conductivity). Thus, for general $p$, the solution should lie somewhere in between these two exact extreme solutions. Also, the approximate expressions should converge to the exact solution for $p$=0 and $p$=1. (Note, substituting $p$=0 in (15) yields (1) with the *max* function in (15) translating to the *min* of the argument of the exponential function in (1)). Second, the proposed solution should converge to the exact solution [18] for the thin wire (i.e. when one of the dimension *a or b* is infinitely large) for general $p$. Third, the expressions should be such that they can be seamlessly applied to the method of spatial resolution that we have discussed in the previous section. The expressions in (15) satisfy these requirements. It is important to note that our approximations neglect 2D interactions due to scattering from orthogonal surfaces, and better approximations accounting for such scattering events can help improve the model in the future. In this paper, we proceed with the expressions in (15) and derive the spatially resolved conductivity for general $p$. Later, we will show how good these approximations are by comparing the average conductivity to the previous works [16], [21].

Following the same process as for $p$=0, we obtain the final SRFS conductivity model as a function of specularity ($p$) as

$$\sigma(x_n, y_n) = \frac{3}{4\pi}\sigma_0 \int_{\theta=0}^{\pi} \eta(x_n, y_n, \theta) cos^2\theta sin\theta d\theta \quad (16)$$

$$\begin{aligned} \eta(x_n, y_n, \theta) &= 2\pi - (1-p) \\ &\times \left[ 2\int_{\Phi=0}^{\frac{\pi}{2}} \frac{\left( e^{\frac{-\kappa_a}{2sin\theta cos\Phi}} \times \cosh\left(\frac{\kappa_a \times x_n}{2sin\theta cos\Phi}\right)\right)}{1-pe^{\frac{-\kappa_a}{sin\theta cos\Phi}}} d\Phi \right. \\ &+ 2\int_{\Phi=0}^{\frac{\pi}{2}} \frac{\left( e^{\frac{-\kappa_b}{2sin\theta sin\Phi}} \times \cosh\left(\frac{\kappa_b \times y_n}{2sin\theta sin\Phi}\right)\right)}{1-pe^{\frac{-\kappa_b}{sin\theta sin\Phi}}} d\Phi \\ &\left. + \frac{1}{2}\sum_{n,d} \int_0^{\frac{\pi}{2}} \left| \frac{e^{\frac{-\kappa_a \times d}{2*sin\theta cos\Phi}}}{1-pe^{\frac{-\kappa_a}{sin\theta cos\Phi}}} - \frac{e^{\frac{-\kappa_b \times n}{2*sin\theta sin\Phi}}}{1-pe^{\frac{-\kappa_b}{sin\theta sin\Phi}}} \right| d\Phi \right] \end{aligned} \quad (17)$$

where $(n,d) \rightarrow \{(1+y_n, 1+x_n), (1-y_n, 1+x_n), (1+y_n, 1-x_n), (1-y_n, 1-x_n)\}$ and $x_n \in [-1,1],\ y_n \in [-1,1]$

The first (' $2\pi$ ') term in (17) corresponds to the bulk conductivity ($\sigma_0$) and the rest of the terms to surface scattering ($\sigma_{SS}$). Note that in the proposed SRFS model, the terms $\sigma_0$ and $\sigma_{SS}$ are subtracted to obtain the overall conductivity. This is different from some recent empirical works [20] which add the resistivity components rather than subtract the conductivity components. Since the subtraction of conductivity terms follows from the rigorous FS treatment (as also shown in [18]), that should be a preferred method to combine different mechanisms as opposed to resistivity addition.

As can be observed from (16) and (17), the SRFS model offers (a) the space-dependence of conductivity as well as (b) its explicit relationship to physical parameters $p$, $\lambda_0$ and the wire cross-section width ($a$) and height ($b$).

*C. Expression for Thin Films*

The expression obtained in (17) is generic and can be applied for thin films of thickness $b$ along the along the $y$-direction by setting $\kappa_a \rightarrow \infty$. The model for a thin film simplifies to

$$\begin{aligned} \frac{\sigma_{SRFS}(y_n)}{\sigma_0} &= \frac{3}{4\pi}\int_{\theta=0}^{\pi} \eta(y_n, \theta) cos^2\theta sin\theta d\theta \\ \eta(y_n, \theta) &= 2\pi - (1-p) \\ &\times 4\int_{\Phi=0}^{\frac{\pi}{2}} \left( \frac{\left( e^{\frac{-\kappa_b}{2sin\theta sin\Phi}} \times \cosh\left(\frac{\kappa_b y_n}{2sin\theta sin\Phi}\right)\right)}{1-p*e^{\frac{-\kappa_b}{sin\theta sin\Phi}}} \right) d\Phi \end{aligned} \quad (18)$$

where $y_n = \frac{y}{b/2} \in [-1,1]$ and $\kappa_b = \frac{b}{\lambda_0}$

## III. Analysis

*A. Proposed SRFS Model and Previous FS-based Models*

As mentioned before, the previous works [13], [22], [23], [24], based on FS models have utilized the simplified versions the *average (not spatially resolved)* conductivity derived in [10] to model the effect of surface scattering on the conductivity of square wires (i.e. $\kappa_a = \kappa_b = \kappa$ ). For different ranges of $\kappa$, different expressions are derived. (It may be noted, however, that several recent works utilize a simple expression for FS conductivity that *must* be used only for cases that $\kappa > 4$, as also pointed out in [16]). Here, we utilize the models in [10] and compare them to the values obtained by averaging the SRFS models across the cross section to validate our approach. The comparison is shown in Fig. 3a illustrating how conductivity increases with increasing $\kappa$. In [10], the expressions assume that all electrons experience inelastic surface scattering ($p$=0). It can be observed in Fig. 3a that the average SRFS model for $p$=0 overlaps with the values in [10], validating the expressions presented in the previous sections. We also show conductivity versus $\kappa$ for other values of $p$. As $p$ increases and electrons experience more elastic scattering, the conductivity rises. For large $p$ and large $\kappa$, the conductivity is close to the bulk value, as expected.

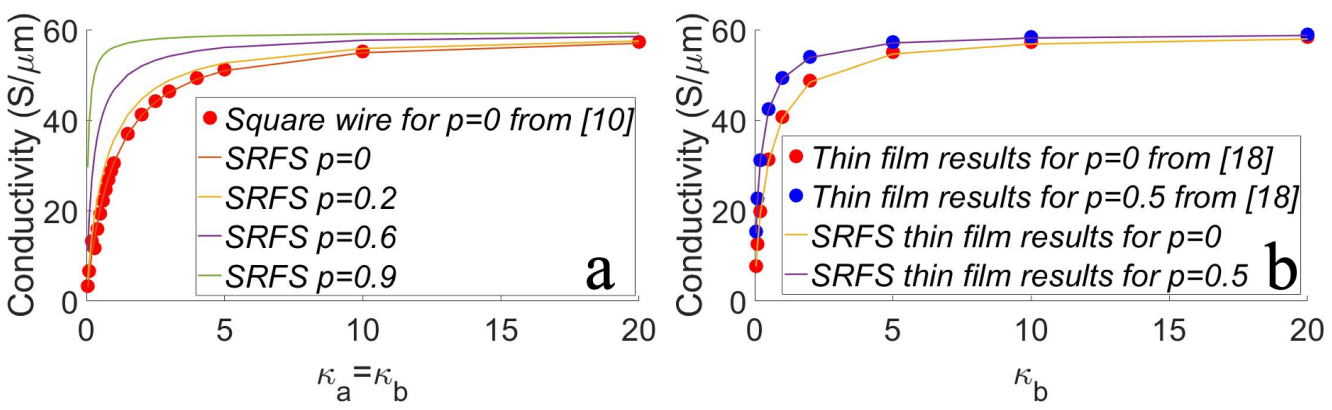

Fig. 3. Conductivity vs $\kappa$ for different specularity $p$ for a) square wire showing the comparison between the proposed SRFS model with the results [10]. b) thin film showing the comparison between the proposed SRFS model with the results [18].

We also compare the FS model with the average conductivity obtained from the proposed SRFS model for a thin film in Fig. 3b. The scatter points are obtained from [18] for $p$ equals to 0 and 0.5. Since our SRFS model is general, we set $\kappa_a$ to be very large value to simulate a thin film and study the effect of different $\kappa_b$ (see Section II C). Our SRFS model matches the results from the original FS model [18] for both $p$ values and wide range of dimensions.

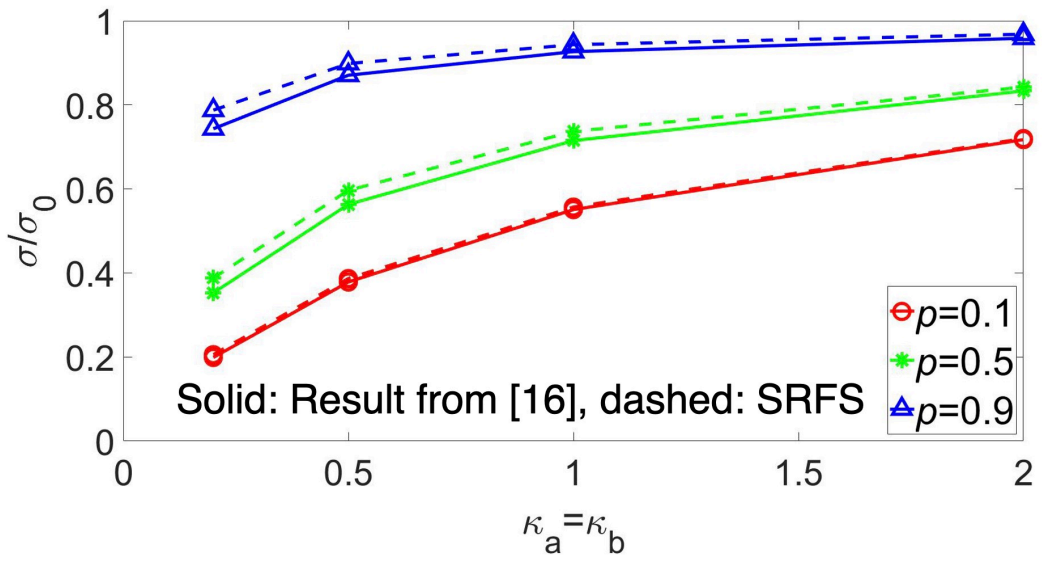

Fig. 4. $\sigma/\sigma_0$ (conductivity normalized to bulk conductivity) for different $\kappa$ and $p$ for a square wire between the approximation proposed (Solid) [16] and SRFS (Dashed), showing SRFS model is able to capture conductivity for general $p$.

*B. Proposed SRFS Model and Kinetic Theory-based Model for Specular Surface Scattering in Square Interconnects*

As noted before, the work in [16] proposes an infinite summation series to model the conductivity *averaged across the cross-section* considering specular surface scattering for rectangular interconnects, as given below:

$$\left(\frac{\sigma}{\sigma_0}\right)_{p,\lambda_0} = (1-p)^2 \sum_{j=1}^{\infty} \left\{ j p^{j-1} \left(\frac{\sigma}{\sigma_0}\right)_{p=0,\lambda_0/j} \right\} \qquad (19)$$

Where the $\frac{\sigma}{\sigma_0}$ ratio for general $p$ is defined as the infinite weighted sum of $\frac{\sigma}{\sigma_0}$ for $p$=0 and $\lambda_0 = \lambda_0/j$. We implement (19) by substituting the average conductivity obtained from our SRFS model for $p$=0 (which, recall, is exact) into the terms $\left(\frac{\sigma}{\sigma_0}\right)_{p=0,\lambda_0/j}$ .We compare the results obtained from this approach to the average conductivity obtained from the proposed SRFS model for general $p$ (see Fig. 4). At small $p$, our SRFS model shows < 3% difference compared to (19). The maximum difference is < 10% when $p$=0.5. As $p$ increases further, the difference reduced again to < 6% @$p$=0.9. Thus, the proposed SRFS approach is able to model the conductivity for general specularity ($p$) in rectangular interconnects with a reasonable accuracy. While the proposed SRFS approach needs to compute the complex integrals only once, the technique from [16] involving infinite sum in (19) needs computation of the complex integrals many times (depending on the maximum value of $j$).

*C. Proposed SRFS Model versus "*cosh*" Model*

The "cosh" model [20] is another spatially resolved conductivity model which uses fitting parameters to match the average conductivity values obtained from the experiment. Here, we compare the spatial dependence of conductivity predicted by the proposed SRFS model with the "cosh" model. We use width of 10nm and thickness of 29nm for the interconnect as used in [20]. Further, following the method in [20], we lump the bulk conductivity and the grain boundary scattering component in the first term of (17) in the SRFS model, and use the same value of this lumped parameter as in [20]. We then match the average conductivity from SRFS model to the "cosh" model by sweeping $p$. Specifically, the "cosh" model reports an average conductivity of 9.8 S/μm (matched to the experiments). We find $p$=0.663 in the SRFS model yielding the same value.

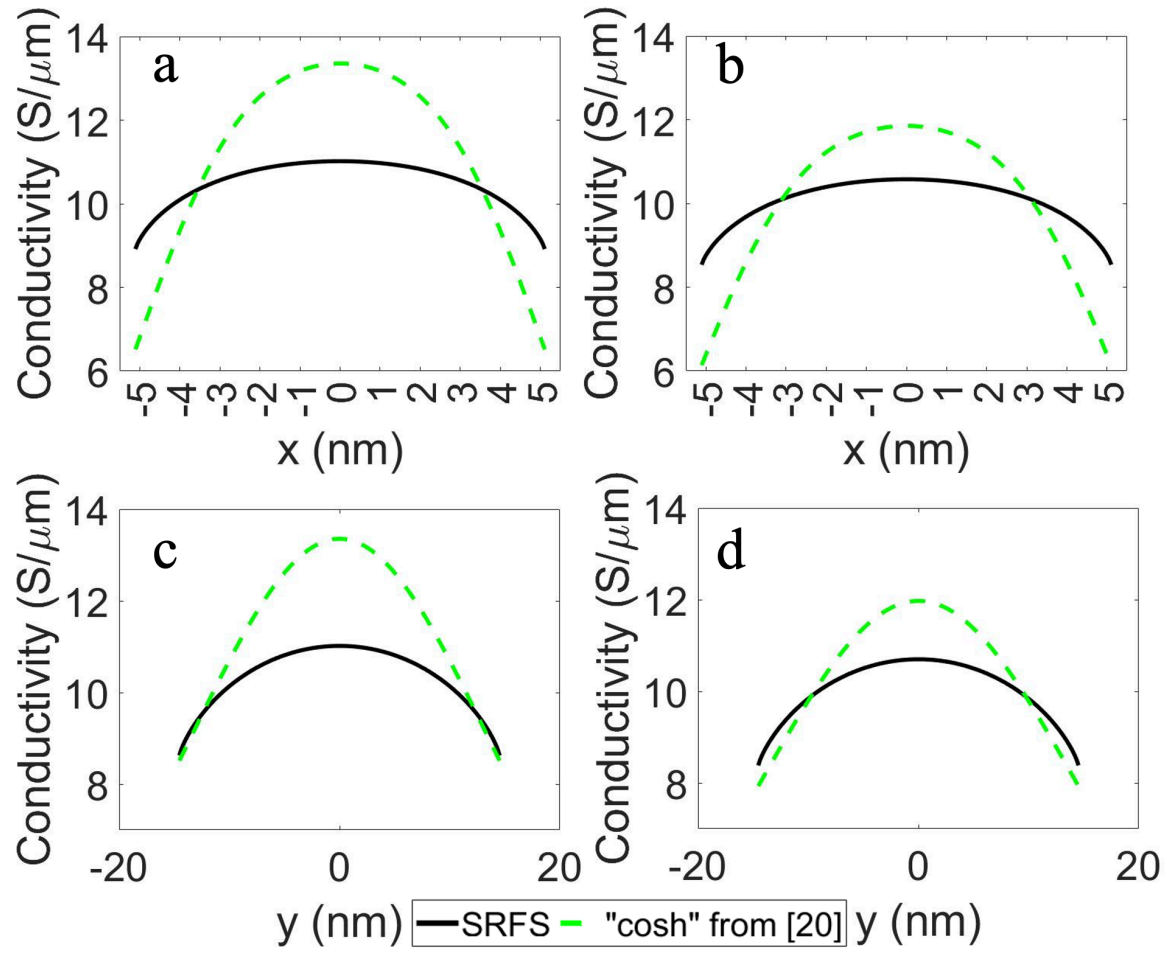

Fig. 5. Conductivity comparison between the proposed SRFS model with "cosh" model along x direction for a) y=0 b) y=7.25nm, along y direction for c) x=0 d) x=2.5nm, highlighting the mismatch between the proposed physical model and an empirical approach.

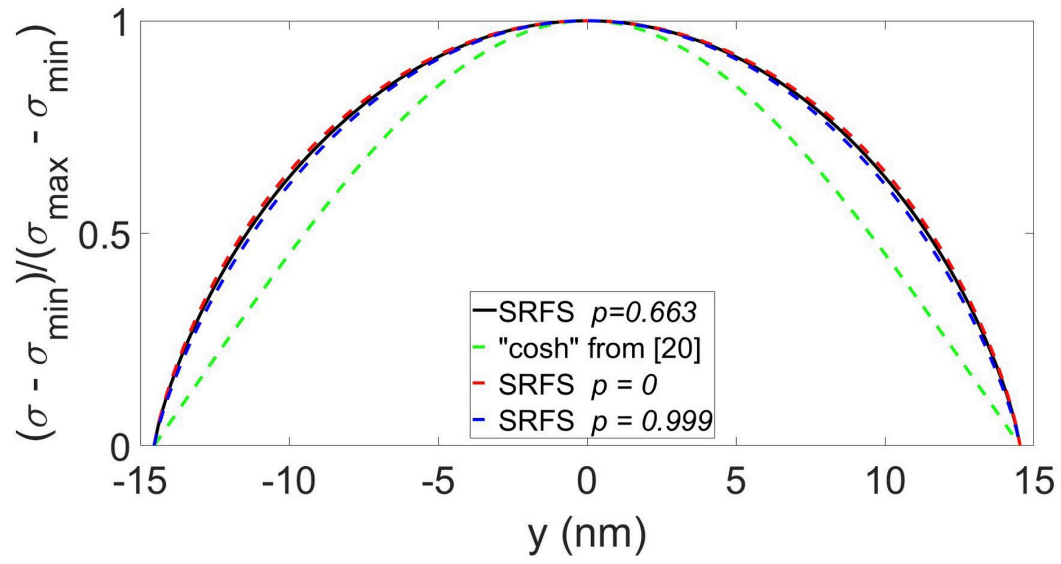

Fig. 6. Spatial profile of normalized conductivity along $y$ direction for $x$=0, listing three different $p$ values for SRFS and "cosh" model [20]. Here, $\sigma_{min}$ and $\sigma_{max}$ are the minimum and maximum $\sigma$. The results show that the SRFS $p$=0.663 spatial profile lies between $p$=0 (exact) and $p\rightarrow 1$.

After matching the average conductivity, we compare the spatial dependence of conductivity obtained from the two models. The conductivity along the width ($x$) direction for y=0 and y=7.25nm are plotted in Fig. 5a and 5b, showing a large mismatch between the physical (SRFS) and the empirical ("cosh") approaches. Similar mismatches are observed for the conductivity along $y$ in Fig. 5c and 5d. The SRFS model has a much flatter conductivity profile compared to the "cosh" model. Although the average conductivity is the same between the two models, the spatial-dependence of conductivity due to surface scattering shows a significant difference, highlighting the importance for a physics-based spatially resolved conductivity model (such as proposed SRFS approach) as opposed to an empirical approach.

Since the proposed SRFS approach for a general $p$ involves some approximations, we show the normalized conductivity profiles for $p$=0 (exact) and $p\rightarrow 1$ along with $p$=0.663 in Fig. 6. The trends illustrate the similar spatial profile for the SRFS model for different values of $p$. As expected, the conductivity spatial profile for p = 0.663 lies between the bounds i.e. $p=0$ and $p\rightarrow 1$. On the other hand, that the spatial profile from "cosh" model lies outside the bounds, suggesting that the proposed SRFS model is more accurate than this empirical approach.

### D. SRFS Conductivity for different κ and p

We show the spatial dependence of conductivity for a square wire predicted by the proposed SRFS model $\kappa=\kappa_a=\kappa_b$=0.2, 1, 5 and $p$=0, 0.9 in Fig. 7. At lower $\kappa$ values, the surface scattering has a larger effect on the conductivity (compared to larger $\kappa$). Thus, the conductivity is smaller at the center. For small $\kappa$, the conductivity reduces smoothly when moving towards the four edges. At larger $\kappa$, the surface scattering effect is reduced at the center; hence the conductivity is closer to bulk conductivity at center. The profile is flat in the center and reduces relatively more sharply closer to the edges (than smaller $\kappa$). With an increase in $p$, the conductivity increases but the shape stays relatively similar.

## IV. Conclusion

We present a 2D spatially resolved FS (SRFS) model for capturing the surface scattering in interconnects with rectangular cross-sections. The primary advantage of our new model is that it offers both space dependence of conductivity and establishes a direct relationship with essential physical parameters. The proposed model is derived from the basic Boltzmann Transport Equation, which relates the spatially resolved conductivity to electron mean free path, specularity, and the dimensions of the conductor. The solution from SRFS is exact for $p$=0 (i.e. diffusive surface scattering). For general $p$ ($p\neq 0$), we make certain approximations and show that the difference for average conductivity across the cross-section between proposed SRFS and previous works [16], [21] is reasonably small. When compared to the 1D [18] and 2D FS models [10], the average conductivity obtained from the proposed model shows a close match. We also show that a previously proposed empirical approach in [20] exhibits a large mismatch in the spatial profile compared to the physics-based modeling approach of SRFS (despite the same average conductivity), highlighting the importance of the latter model presented in this work.

The rigorous nature of the SRFS model leads to complex expressions. While these expressions are easy to implement in standard software such as MATLAB, they may be too time consuming to solve in a circuit simulator. Therefore, building on the SRFS model presented in this part, we present a compact model for spatially resolved conductivity accounting for surface scattering in rectangular interconnects in part II of this work.

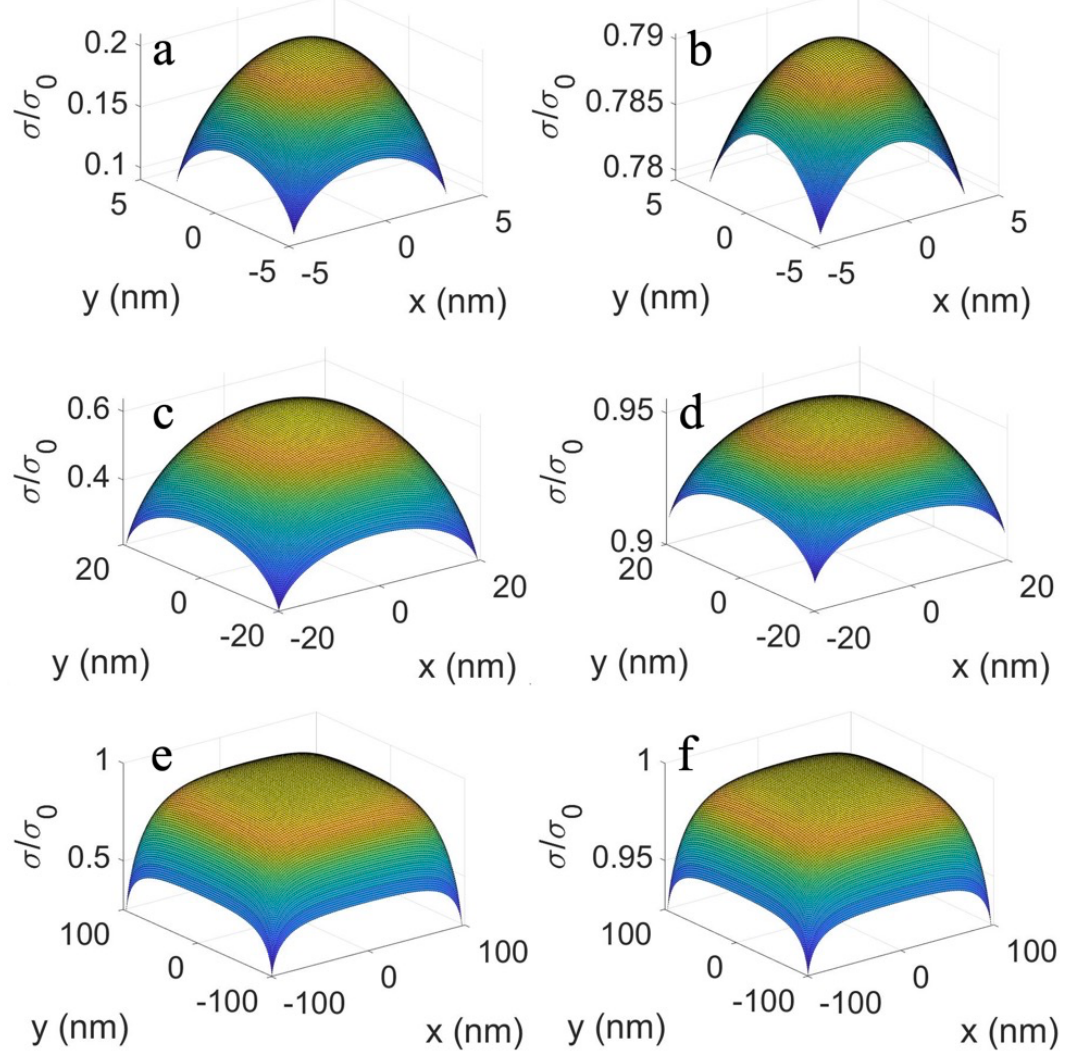

Fig. 7. The spatial profiles for $\sigma/\sigma_0$ (conductivity normalized to bulk conductivity) for different $\kappa$ and $p$ for a square wire, a) *κ=0.2 p=0 b) κ=0.2, p=0.9 c) κ=1, p=0 d) κ=1, p=0.9, e) κ=5, p=0 f) κ=5, p=0.9* showing as $p$ increases conductivity also increases, as $\kappa$ increases conductivity increases to bulk conductivity especially at the center of the wire.

## Appendix A1

When we integrate (12), in general, there are eight different cases (depending on the spatial location) for the order of the four boundary points in the range from 0 to $\frac{\pi}{2}$. However, we solve (12) following an approach which makes the order of the boundary points irrelevant.

To illustrate this approach, we take a region where $x_n$<0 and $y_n$<$x_n$ as an example. For this region, $\eta$ function can be written as (a1)

$$\eta(x_n, y_n, \theta) = \tag{a1}$$

$$\int_0^{tan^{-1}\left(\frac{b}{a}\left(\frac{1+y_n}{1-x_n}\right)\right)} \left\{ \sum_{\substack{\alpha_i=\alpha_{x+},\alpha_{x-},\\ \alpha_{x+},\alpha_{x-}}} \xi(\alpha_i) \right\} d\Phi + \int_{tan^{-1}\left(\frac{b}{a}\left(\frac{1+y_n}{1-x_n}\right)\right)}^{tan^{-1}\left(\frac{b}{a}\left(\frac{1+y_n}{1+x_n}\right)\right)} \left\{ \sum_{\substack{\alpha_i=\alpha_{x+},\alpha_{y+},\\ \alpha_{x+},\alpha_{x-}}} \xi(\alpha_i) \right\} d\Phi$$

$$+ \int_{tan^{-1}\left(\frac{b}{a}\left(\frac{1+y_n}{1+x_n}\right)\right)}^{tan^{-1}\left(\frac{b}{a}\left(\frac{1-y_n}{1-x_n}\right)\right)} \left\{ \sum_{\substack{\alpha_i=\alpha_{y+},\alpha_{y+},\\ \alpha_{x+},\alpha_{x-}}} \xi(\alpha_i) \right\} d\Phi + \int_{tan^{-1}\left(\frac{b}{a}\left(\frac{1-y_n}{1-x_n}\right)\right)}^{tan^{-1}\left(\frac{b}{a}\left(\frac{1-y_n}{1+x_n}\right)\right)} \left\{ \sum_{\substack{\alpha_i=\alpha_{y+},\alpha_{y+},\\ \alpha_{x+},\alpha_{y-}}} \xi(\alpha_i) \right\} d\Phi$$

$$+ \int_{tan^{-1}\left(\frac{b}{a}\left(\frac{1-y_n}{1+x_n}\right)\right)}^{\frac{\pi}{2}} \left\{ \sum_{\substack{\alpha_i=\alpha_{y+},\alpha_{y+},\\ \alpha_{y-},\alpha_{y-}}} \xi(\alpha_i) \right\} d\Phi$$

Here $\xi(\alpha_i) = 1 - e^{\frac{-\alpha_i}{2sin\theta}}$ and $\alpha_{x+} = \frac{\kappa_a(1+x_n)}{cos\Phi}, \alpha_{x-} = \frac{\kappa_a(1-x_n)}{cos\Phi}, \alpha_{y+} = \frac{\kappa_b(1+y_n)}{sin\Phi}, \alpha_{y-} = \frac{\kappa_b(1-y_n)}{sin\Phi}$. Equation (a1) can be rewritten as follows by changing the upper limits of all the integrals to $\frac{\pi}{2}$.

$$\eta(x_n, y_n, \theta) = \tag{a2}$$

$$\int_0^{\frac{\pi}{2}} \left\{ \sum_{\substack{\alpha_i=\alpha_{x+},\alpha_{x-},\\ \alpha_{x+},\alpha_{x-}}} \xi(\alpha_i) \right\} d\Phi - \int_{tan^{-1}\left(\frac{b}{a}\left(\frac{1+y_n}{1-x_n}\right)\right)}^{\frac{\pi}{2}} \left\{ \sum_{\substack{\alpha_i=\alpha_{x+},\alpha_{x-},\\ \alpha_{x+},\alpha_{x-}}} \xi(\alpha_i) \right\} d\Phi$$

$$+ \int_{tan^{-1}\left(\frac{b}{a}\left(\frac{1+y_n}{1-x_n}\right)\right)}^{\frac{\pi}{2}} \left\{ \sum_{\substack{\alpha_i=\alpha_{x+},\alpha_{y+},\\ \alpha_{x+},\alpha_{x-}}} \xi(\alpha_i) \right\} d\Phi - \int_{tan^{-1}\left(\frac{b}{a}\left(\frac{1+y_n}{1+x_n}\right)\right)}^{\frac{\pi}{2}} \left\{ \sum_{\substack{\alpha_i=\alpha_{x+},\alpha_{y+},\\ \alpha_{x+},\alpha_{x-}}} \xi(\alpha_i) \right\} d\Phi$$

$$+ \int_{tan^{-1}\left(\frac{b}{a}\left(\frac{1+y_n}{1+x_n}\right)\right)}^{\frac{\pi}{2}} \left\{ \sum_{\substack{\alpha_i=\alpha_{y+},\alpha_{y+},\\ \alpha_{x+},\alpha_{x-}}} \xi(\alpha_i) \right\} d\Phi - \int_{tan^{-1}\left(\frac{b}{a}\left(\frac{1-y_n}{1-x_n}\right)\right)}^{\frac{\pi}{2}} \left\{ \sum_{\substack{\alpha_i=\alpha_{y+},\alpha_{y+},\\ \alpha_{x+},\alpha_{x-}}} \xi(\alpha_i) \right\} d\Phi$$

$$+ \int_{tan^{-1}\left(\frac{b}{a}\left(\frac{1-y_n}{1-x_n}\right)\right)}^{\frac{\pi}{2}} \left\{ \sum_{\substack{\alpha_i=\alpha_{y+},\alpha_{y+},\\ \alpha_{x+},\alpha_{y-}}} \xi(\alpha_i) \right\} d\Phi - \int_{tan^{-1}\left(\frac{b}{a}\left(\frac{1-y_n}{1+x_n}\right)\right)}^{\frac{\pi}{2}} \left\{ \sum_{\substack{\alpha_i=\alpha_{y+},\alpha_{y+},\\ \alpha_{x+},\alpha_{y-}}} \xi(\alpha_i) \right\} d\Phi$$

$$+ \int_{tan^{-1}\left(\frac{b}{a}\left(\frac{1-y_n}{1+x_n}\right)\right)}^{\frac{\pi}{2}} \left\{ \sum_{\substack{\alpha_i=\alpha_{y+},\alpha_{y+},\\ \alpha_{y-},\alpha_{y-}}} \xi(\alpha_i) \right\} d\Phi$$

If we observe the second and third term from (a2), they have the same integration limits and differ from each other only in one of the four sub-expressions dependent on $\alpha_i$. By summing the second and third expressions, we obtain

$$\int_{tan^{-1}\left(\frac{b}{a}\left(\frac{1+y_n}{1-x_n}\right)\right)}^{\frac{\pi}{2}} \left(\xi(\alpha_{y+}) - \xi(\alpha_{x-})\right) d\Phi \tag{a3}$$

We repeat this for all the terms #2 through #9. Also, the first term in (a2) is rewritten as

$$4\int_{\Phi=0}^{\frac{\pi}{2}} d\Phi - 4\int_{\Phi=0}^{\frac{\pi}{2}} \left(e^{\frac{-\kappa_a}{2sin\theta cos\Phi}} \times \cosh\left(\frac{\kappa_a \times x_n}{2sin\theta cos\Phi}\right)\right) d\Phi \tag{a4}$$

Thus, the $\eta$ function can be simplified as follow

$$\eta(x_n, y_n, \theta) = 4\int_{\Phi=0}^{\frac{\pi}{2}} d\Phi - 4\int_{\Phi=0}^{\frac{\pi}{2}} \left(e^{\frac{-\kappa_a}{2sin\theta cos\Phi}} \times \cosh\left(\frac{\kappa_a \times x_n}{2sin\theta cos\Phi}\right)\right) d\Phi \tag{a5}$$

$$- \left\{ \int_{tan^{-1}\left(\frac{b}{a}\left(\frac{1+y_n}{1-x_n}\right)\right)}^{\frac{\pi}{2}} \left(\xi(\alpha_{y+}) - \xi(\alpha_{x-})\right) d\Phi + \int_{tan^{-1}\left(\frac{b}{a}\left(\frac{1+y_n}{1+x_n}\right)\right)}^{\frac{\pi}{2}} \left(\xi(\alpha_{y+}) - \xi(\alpha_{x+})\right) d\Phi \right.$$

$$\left. + \int_{tan^{-1}\left(\frac{b}{a}\left(\frac{1-y_n}{1-x_n}\right)\right)}^{\frac{\pi}{2}} \left(\xi(\alpha_{y-}) - \xi(\alpha_{x-})\right) d\Phi + \int_{tan^{-1}\left(\frac{b}{a}\left(\frac{1-y_n}{1+x_n}\right)\right)}^{\frac{\pi}{2}} \left(\xi(\alpha_{y-}) - \xi(\alpha_{x+})\right) d\Phi \right\}$$

If we look at the terms within the curly bracket, the order of the boundary points does not matter. The expression is only a function of the boundary points. Thus, this $\eta$ expression holds true for any location of the interconnect cross section (not just the region considered before).

Recall, when we transformed (a1) to (a2), we make the upper limits of all the integrals to be $\frac{\pi}{2}$ and then combine the expressions. We can also follow a similar process but make the lower limit of the integrals to be 0. By completing the same steps, we can get another expression for $\eta$ function (a6).

$$\eta(x_n, y_n, \theta) = 4\int_{\Phi=0}^{\frac{\pi}{2}} d\Phi - 4\int_{\Phi=0}^{\frac{\pi}{2}} \left(e^{\frac{-\kappa_b}{2sin\theta sin\Phi}} \times \cosh\left(\frac{\kappa_b \times y_n}{2sin\theta sin\Phi}\right)\right) d\Phi \tag{a6}$$

$$- \left\{ \int_0^{tan^{-1}\left(\frac{b}{a}\left(\frac{1-y_n}{1+x_n}\right)\right)} \left(\xi(\alpha_{x+}) - \xi(\alpha_{y-})\right) d\Phi + \int_0^{tan^{-1}\left(\frac{b}{a}\left(\frac{1-y_n}{1-x_n}\right)\right)} \left(\xi(\alpha_{x-}) - \xi(\alpha_{y-})\right) d\Phi \right.$$

$$\left. + \int_0^{tan^{-1}\left(\frac{b}{a}\left(\frac{1+y_n}{1+x_n}\right)\right)} \left(\xi(\alpha_{x+}) - \xi(\alpha_{y+})\right) d\Phi + \int_0^{tan^{-1}\left(\frac{b}{a}\left(\frac{1+y_n}{1-x_n}\right)\right)} \left(\xi(\alpha_{x-}) - \xi(\alpha_{y+})\right) d\Phi \right\}$$

Now, we add the two $\eta$ functions in (a5) and (a6) together and divide by 2 to get the final expression (14). Note, when we combine two $\eta$ functions (a5) and (a6), we add the two integrals corresponding to the same boundary points in the upper/lower limit of integration to obtain the final integral with $\Phi = 0 \rightarrow \pi/2$. For instance, we add the first term in the curly braces in (a5) with the last term in the curly braces in (a6) and obtain

$$\int_{tan^{-1}\left(\frac{b}{a}\left(\frac{1+y_n}{1-x_n}\right)\right)}^{\frac{\pi}{2}} \left(\xi(\alpha_{y+}) - \xi(\alpha_{x-})\right) d\Phi$$

$$+ \int_0^{tan^{-1}\left(\frac{b}{a}\left(\frac{1+y_n}{1-x_n}\right)\right)} \left(\xi(\alpha_{x-}) - \xi(\alpha_{y+})\right) d\Phi \tag{a7}$$

$$= \int_0^{\frac{\pi}{2}} \left|\xi(\alpha_{x-}) - \xi(\alpha_{y+})\right| d\Phi$$

It is noteworthy that the two integrals on the left hand side of (a7) can be combined into one (right hand side) as the corresponding integrands are the same except for a negative sign. Also, the integrands are always positive, as dictated by the *min* function in (1). Hence, we use the absolute value of $\xi(\alpha_{x-}) - \xi(\alpha_{y+})$ on the right hand side of (a7). Similarly we combine the other integrals in (a5) and (a6) to obtain the final expression in (14). This approach is also valid the approximate expressions obtained for a general *p*.

## REFERENCES

[1] R. Brain, "Interconnect scaling: Challenges and opportunities," in *2016 IEEE International Electron Devices Meeting (IEDM)*, 2016, pp. 9.3.1-9.3.4. doi: 10.1109/IEDM.2016.7838381.

[2] R. L. Graham *et al.*, "Resistivity dominated by surface scattering in sub-50 nm Cu wires," *Appl Phys Lett*, vol. 96, no. 4, 2010, doi: 10.1063/1.3292022.

[3] W. L. Wang *et al.*, "The Reliability Improvement of Cu Interconnection by the Control of Crystallized α-Ta/TaNx Diffusion Barrier," *J Nanomater*, vol. 2015, 2015, doi: 10.1155/2015/917935.

[4] D. Edelstein *et al.*, "A high performance liner for copper damascene interconnects," in *Proceedings of the IEEE 2001 International Interconnect Technology Conference (Cat. No.01EX461)*, 2001, pp. 9–11. doi: 10.1109/IITC.2001.930001.

[5] N. Bekiaris *et al.*, "Cobalt fill for advanced interconnects," in *2017 IEEE International Interconnect Technology Conference (IITC)*, 2017, pp. 1–3. doi: 10.1109/IITC-AMC.2017.7968981.

[6] D. Wan *et al.*, "Subtractive Etch of Ruthenium for Sub-5nm Interconnect," in *2018 IEEE International Interconnect Technology Conference (IITC)*, 2018, pp. 10–12. doi: 10.1109/IITC.2018.8454841.

[7] C. L. Lo *et al.*, "Opportunities and challenges of 2D materials in back-end-of-line interconnect scaling," Aug. 28, 2020, *American Institute of Physics Inc.* doi: 10.1063/5.0013737.

[8] B. Grot, J. Hestness, S. W. Keckler, and O. Mutlu, "Express Cube Topologies for on-Chip Interconnects," in *2009 IEEE 15th International Symposium on High Performance Computer Architecture*, 2009, pp. 163–174. doi: 10.1109/HPCA.2009.4798251.

[9] E. H. SONDHEIMER, "Influence of a Magnetic Field on the Conductivity of Thin Metallic Films," *Nature*, vol. 164, no. 4178, pp. 920–921, 1949, doi: 10.1038/164920a0.

[10] D. K. C. Macdonald and K. Sarginson, "Size Effect Variation of the Electrical Conductivity of Metals," 1950. [Online]. Available: https://about.jstor.org/terms

[11] R. B. Dingle and W. L. Bragg, "The electrical conductivity of thin wires," *Proc R Soc Lond A Math Phys Sci*, vol. 201, no. 1067, pp. 545–560, 1950, doi: 10.1098/rspa.1950.0077.

[12] A. F. Mayadas and M. Shatzkes, "Electrical-Resistivity Model for Polycrystalline Films: the Case of Arbitrary Reflection at External Surfaces," *Phys Rev B*, vol. 1, no. 4, pp. 1382–1389, Feb. 1970, doi: 10.1103/PhysRevB.1.1382.

[13] R. Mehta, S. Chugh, and Z. Chen, “Enhanced electrical and thermal conduction in graphene-encapsulated copper nanowires,” *Nano Lett*, vol. 15, no. 3, pp. 2024–2030, Mar. 2015, doi: 10.1021/nl504889t.

[14] T. Shen *et al.*, “MoS2 for Enhanced Electrical Performance of Ultrathin Copper Films,” *ACS Appl Mater Interfaces*, vol. 11, no. 31, pp. 28345–28351, Aug. 2019, doi: 10.1021/acsami.9b03381.

[15] X. Chen, C.-L. Lo, M. C. Johnson, Z. Chen, and S. K. Gupta, “Modeling and Circuit Analysis of Interconnects with TaS2 Barrier/Liner,” in *2021 Device Research Conference (DRC)*, 2021, pp. 1–2. doi: 10.1109/DRC52342.2021.9467160.

[16] R. G. Chambers, “The Conductivity of Thin Wires in a Magnetic Field,” 1950. [Online]. Available: https://www.jstor.org/stable/98525

[17] L. Moraga, C. Arenas, R. Henriquez, and B. Solis, “The effect of surface roughness and grain-boundary scattering on the electrical conductivity of thin metallic wires,” *Phys Status Solidi B Basic Res*, vol. 252, no. 1, pp. 219–229, Jan. 2015, doi: 10.1002/pssb.201451202.

[18] E. H. Sondheimer, “The mean free path of electrons in metals,” *Adv Phys*, vol. 1, no. 1, pp. 1–42, 1952, doi: 10.1080/00018735200101151.

[19] T. Lu and A. Srivastava, “Detailed electrical and reliability study of tapered TSVs,” in *2013 IEEE International 3D Systems Integration Conference (3DIC)*, 2013, pp. 1–7. doi: 10.1109/3DIC.2013.6702350.

[20] I. Ciofi *et al.*, “Impact of Wire Geometry on Interconnect RC and Circuit Delay,” *IEEE Trans Electron Devices*, vol. 63, no. 6, pp. 2488–2496, Jun. 2016, doi: 10.1109/TED.2016.2554561.

[21] W. Steinhögl, G. Schindler, G. Steinlesberger, and M. Engelhardt, “Size-dependent resistivity of metallic wires in the mesoscopic range,” *Phys Rev B Condens Matter Mater Phys*, vol. 66, no. 7, pp. 1–4, 2002, doi: 10.1103/PhysRevB.66.075414.

[22] N. C. Wang, S. Sinha, B. Cline, C. D. English, G. Yeric, and E. Pop, “Replacing copper interconnects with graphene at a 7-nm node,” in *2017 IEEE International Interconnect Technology Conference (IITC)*, 2017, pp. 1–3. doi: 10.1109/IITC-AMC.2017.7968949.

[23] R. Saligram, S. Datta, and A. Raychowdhury, “Scaled Back End of Line Interconnects at Cryogenic Temperatures,” *IEEE Electron Device Letters*, vol. 42, no. 11, pp. 1674–1677, Nov. 2021, doi: 10.1109/LED.2021.3117277.

[24] A. Pyzyna *et al.*, “Resistivity of copper interconnects beyond the 7 nm node,” 2015. [Online]. Available: http://www.itrs.net